\documentclass[pra,showpacs,twocolumn]{revtex4}
\usepackage{psfrag}
\usepackage{graphicx}

\begin{document}

\title{Classical Correlation and Quantum Discord in Critical Systems}
\author{M. S. \surname{Sarandy}}
\email{msarandy@if.uff.br}
\affiliation{Instituto de F\'{\i}sica, Universidade Federal Fluminense,
Av. Gal. Milton Tavares de Souza s/n, Gragoat\'a, 24210-346, Niter\'oi, RJ, Brazil.}
\date{\today }

\begin{abstract}
We discuss the behavior of quantum and classical pairwise correlations 
in critical systems, with the quantumness of the correlations measured 
by the quantum discord. We analytically derive these correlations for general 
real density matrices displaying $Z_2$ symmetry. As an illustration, we  
analyze both the XXZ and the transverse field Ising models. Finite-size as well 
as infinite chains are investigated and the quantum criticality is discussed. 
Moreover, we identify the spin functions that govern the correlations. 
As a further example, we also consider correlations in the Hartree-Fock 
ground state of the Lipkin-Meshkov-Glick model. It is then shown that both classical 
correlation and quantum discord exhibit signatures of the quantum phase transitions.
\end{abstract}


\maketitle

%%%%%%%%%%%%%%%%%%%%%%%%%%%%%%%%%%%%%%%%%%%%%%%%%%%%%
\section{Introduction}
%%%%%%%%%%%%%%%%%%%%%%%%%%%%%%%%%%%%%%%%%%%%%%%%%%%%%

\label{introduction}

The concept of correlation, i.e., information of one system 
about another, is a key element in many-body physics. Indeed, 
the properties of a many-body system are strongly 
affected by changes in the correlations among its constituents. 
These changes are responsible for the occurrence of remarkable
phenomena such as a quantum phase transition (QPT), which is a 
critical change in the ground state of a 
quantum system due to level crossings in its energy spectrum. 
QPTs occur at low temperatures $T$ where the de Broglie  
wavelength is greater than the correlation length of 
the thermal fluctuations (effectively $T=0$)~\cite{Sachdev:book,Continentino:book}. 
Striking examples of QPTs include metal-insulator transitions in strongly correlated 
electronic materials, magnetic transitions in quantum spin lattices, superfluid-Mott insulator 
transitions in atomic gases induced by a Bose-Einstein condensation, among others.

Correlations can be both from classical and quantum sources. The existence of 
genuinely quantum correlations can be usually inferred by the presence of 
entanglement among parts of a system. Indeed, entanglement displays a rather interesting 
behavior at QPTs~\cite{Amico:08}, being able to indicate a quantum critical point (QCP)  
through nonanalyticities inherited from the ground state energy~\cite{Osterloh:02,Wu:04}. 
Moreover, for one-dimensional critical systems, ground state entanglement entropy 
exhibits a universal logarithmic scaling governed by the central charge of the Virasoro 
Algebra associated with the underlying conformal field theory~\cite{Vidal:03,Korepin:04,Calabrese:04}. 
Remarkably, the logarithmic scaling is robust against disorder~\cite{Refael:04,Saguia:07,Lin:07,Hur:07}. 
In higher dimensions, entanglement usually scales following an area law for noncritical 
systems (see, e.g., Ref.~\cite{Plenio:05}). For critical models, violations of the area law have 
been found~\cite{Wolf:06,Yu:08}, with logarithmic-type corrections appearing. More recently, 
it has been observed that an area law is generally implied by a finite correlation length when measured 
in terms of the mutual information~\cite{Wolf:08}.  
This remarkable behavior of entanglement at criticality is indeed a consequence of the   
correlation pattern exhibited by the ground state of the system. 

Nevertheless, although entanglement provides a route to find out the existence of quantum 
correlations, it can be shown that they can appear even when entanglement 
is absent~\cite{Ollivier:02,Luo:08}. Quantum correlations, which can be measured by 
the {\it quantum discord}~\cite{Ollivier:02}, often arise as a consequence of coherence 
in a quantum system, being present even for separable states. Moreover, entangled states 
commonly involve more than only quantum correlations, i.e., they usually carry classical correlations 
among their parts. 
%%%% V2: sentence below has been rewritten and a new ref. has been included.
In a multiparty mixed-state scenario, the possibility of nonseparable states with purely 
quantum correlations (with no supporting background of classical correlations) have also 
been investigated~\cite{Kaszlikowski:08}, but genuine indicators of multiparticle classical 
correlations are still under debate~\cite{Horodecki:08}. 
%%%%%%
By focusing on condensed-matter systems, the aim of this work is, starting from an arbitrary 
$Z_2$-symmetric model, to investigate pairwise correlations, explicitly splitting 
up their classical and quantum contributions and analyzing their behavior at QPTs. As an 
illustration, ground states of both XXZ and transverse field Ising spin chains will be shown to share 
quantum as well as classical correlations for nearest-neighbor spin pairs, with both of them signalling 
the critical behavior of the system. An analysis of such correlations for the thermodynamic 
limit of XXZ and Ising chains has recently appeared in Ref.~\cite{Dillenschneider:08}. Here, 
we generalize this analysis by providing analytical expressions for the correlations 
in each model and also by discussing their finite-size behavior. As a further illustration, 
we describe the correlations in the Hartree-Fock ground state of many-body systems, taking the 
Lipkin-Meshkov-Glick (LMG) model as an example. 

The paper is organized as follows. In Sec.~\ref{sec-qd} we review the evaluation of correlations from the 
point of view of information theory, describing in particular the concept of quantum discord. In Sec.~\ref{sec-sl}, 
we discuss the computation of classical and quantum correlations for general models 
displaying $Z_2$ symmetry. Sec.~\ref{sec-xxz}, Sec.~\ref{sec-ib}, and Sec.~\ref{sec-lip} are devoted to 
illustrating our results for the XXZ chain, the transverse field Ising chain, and the LMG model, respectively. 
Conclusions are then presented in Sec.~\ref{sec-con}.

%%%%%%%%%%%%%%%%%%%%%%%%%%%%%%%%%%%%%%%%%%%%%%%%%%%%%%%%%%%%%%%%%%%%%%%%%%%%%%%%%%%%
\section {Classical correlations and quantum discord}
%%%%%%%%%%%%%%%%%%%%%%%%%%%%%%%%%%%%%%%%%%%%%%%%%%%%%%%%%%%%%%%%%%%%%%%%%%%%%%%%%%%%

\label{sec-qd}

In classical information theory, the information obtained, on average, after knowing 
the value of a random variable $X$, which takes values within a set of probabilities 
$\{p_x\}$,  can be quantified by its Shannon entropy $H(X) = - \sum_x p_x \log p_x$. 
We use the symbol $\log$ as denoting logarithm at base 2 throughout the text.
By taking two such random variables $X$ and $Y$, we can measure the correlation between 
them by their mutual information 
\begin{equation}
{\cal I}(X:Y) = H(X) + H(Y) - H(X,Y),
\label{mi1}
\end{equation}
where $H(X,Y) = - \sum_{x,y} p_{xy} \log p_{xy}$ is the joint entropy for $X$ and $Y$. 
By introducing the conditional entropy 
\begin{equation}
H(X|Y) = H(X,Y) - H(Y), 
\label{cce1}
\end{equation}
which quantifies the ignorance (on average) about the value of X given Y is known, we 
can rewrite Eq.~(\ref{mi1}) as 
\begin{equation}
{\cal I}(X:Y) = H(X) - H(X|Y).
\label{mi2}
\end{equation}
In order to generalize the above equations to the quantum domain, we replace classical 
probability distributions by density matrices. Denoting by $\rho$ the density matrix of a 
composite system $AB$ and by $\rho^A$ and $\rho^B$ the density matrices of parts $A$ and $B$, 
respectively, the quantum mutual information can be defined as
\begin{equation}
I(\rho^A : \rho^B) = S(\rho^A) - S(\rho^A | \rho^B),
\label{qmi1}
\end{equation}
where $S(\rho^A) = -{\textrm{Tr}} \rho^A \log \rho^A$ is the von Neumann entropy for subsystem $A$ and
\begin{equation} 
S(\rho^A | \rho^B) = S(\rho) - S(\rho^B)
\label{qce1}
\end{equation}
is a quantum generalization of the conditional entropy for $A$ and $B$. A remarkable observation 
realized in Ref.~\cite{Ollivier:02} is that the conditional entropy can be introduced by a different 
approach which, although classically equivalent to Eq.~(\ref{cce1}), yields a result in the quantum 
case that differs from Eq.~(\ref{qce1}). Indeed, let us consider a measurement performed 
locally only on part $B$. This measurement can be described by a set of projectors $\{B_k\}$. 
The state of the quantum system, conditioned on the measurement of the outcome labelled by $k$, 
becomes
\begin{equation}
\rho_k = \frac{1}{p_k} \left( I\otimes B_k \right) \rho \left( I\otimes B_k \right), 
\label{rhok}
\end{equation}
where $p_k = {\textrm{Tr}} [ (I\otimes B_k ) \rho ( I\otimes B_k ) ]$ denotes the probability of 
obtaining the outcome $k$ and $I$ denotes the identity operator for the subsystem $A$. The conditional 
density operator given by Eq.~(\ref{rhok}) allows for the following alternative definition of the quantum conditional entropy:
\begin{equation}
S(\rho|\{B_k\}) = \sum_k p_k S(\rho_k).
\end{equation}
Therefore, following Eq.~(\ref{mi2}), the quantum mutual information can also be alternatively defined by
\begin{equation}
J(\rho:\{B_k\}) = S(\rho^A) - S(\rho|\{B_k\}).
\label{qmi2}
\end{equation}
Eqs.~(\ref{qmi1}) and~(\ref{qmi2}) are classically equivalent but they are different in the quantum case. The 
difference between them is due to quantum effects on the correlation between parts $A$ and $B$ and provides  
a measure for the quantumness of the correlation, which has been called {\it quantum discord}~\cite{Ollivier:02}.  
In fact, following Refs.~\cite{Ollivier:02,Henderson:01}, we can define the classical correlation between 
parts $A$ and $B$ as
\begin{equation}
C(\rho) = \max_{\{B_k\}} J(\rho:\{B_k\}),
\label{ccorrel}
\end{equation}
with the quantum correlation accounted by the quantum discord, which is then given by 
\begin{equation}
Q(\rho) = I(\rho^A : \rho^B) -  C(\rho).
\label{qcorrel}
\end{equation}
      
%%%%%%%%%%%%%%%%%%%%%%%%%%%%%%%%%%%%%%%%%%%%%%%%%%%%%%%%%%%%%%%%%%%%%%%%%%%%%%%%%%%%
\section {Pairwise correlations for $Z_2$-symmetric quantum spin lattices}
%%%%%%%%%%%%%%%%%%%%%%%%%%%%%%%%%%%%%%%%%%%%%%%%%%%%%%%%%%%%%%%%%%%%%%%%%%%%%%%%%%%%

\label{sec-sl}

We will consider here an interacting pair of spins-1/2 in a spin lattice, 
which is governed by a Hamiltonian $H$ that is both real and exhibits $Z_2$ symmetry, 
i.e. invariance under $\pi$-rotation around a given spin axis. By taking this spin axis as the 
$z$ direction, this implies the commutation of $H$ with the  parity operator 
$\bigotimes_{i=1}^{N} \sigma^3_i$, where $N$ denotes the total number of spins and 
$\sigma^3_i$ is the Pauli operator along the $z$-axis at site $i$. Note that a number of spin 
models are enclosed within these requirements as, for instance, the XXZ spin chain and the 
transverse field Ising model. Disregarding spontaneous symmetry breaking (see, e.g., 
%%%%%%%%%%% V2: One Ref included below
Refs.~\cite{Syljuasen:03,Osterloh:06,Oliveira:08} 
%%%%%%%%%%%%%%%%%%%%%%%%%%%%
for a treatment of spontaneously broken ground states), 
the two-spin reduced density matrix at sites labelled by $i$ and $j$ in the basis 
$\{ |\uparrow\uparrow\rangle, |\uparrow\downarrow\rangle, |\downarrow\uparrow\rangle, |\downarrow\downarrow\rangle\}$, 
with $|\uparrow\rangle$ and $|\downarrow\rangle$ denoting the eigenstates of  $\sigma^3$, will be given by
\begin{equation}
\mathcal{\rho}=\left( 
\begin{array}{cccc}
a & 0 & 0 & f \\ 
0 & b_1 & z & 0 \\ 
0 & z & b_2 & 0 \\ 
f & 0 & 0 & d
\end{array}
\right) .  \label{rhoAB}
\end{equation}
In terms of spin correlation functions, these elements can be written as 
\begin{eqnarray}
a &=& \frac{1}{4} \left(1+G^i_{z}+G^{j}_z+G^{ij}_{zz}\right) \, , \nonumber \\
b_1 &=& \frac{1}{4} \left(1+G^i_{z}-G^{j}_z-G^{ij}_{zz}\right) \, , \nonumber \\
b_2 &=& \frac{1}{4} \left(1-G^i_{z}+G^{j}_z-G^{ij}_{zz}\right) \, , \nonumber \\
d &=& \frac{1}{4} \left(1-G^i_{z}-G^{j}_z+G^{ij}_{zz}\right) \, , \nonumber \\
z &=& \frac{1}{4} \left(G^{ij}_{xx}+G^{ij}_{yy} \right),  \nonumber \\
f &=& \frac{1}{4} \left(G^{ij}_{xx}-G^{ij}_{yy} \right), 
\label{relem}
\end{eqnarray}
where $G^k_{z} = \langle \sigma_z^k \rangle$ $(k=i,j)$ is the magnetization density at site $k$ and 
$G^{ij}_{\alpha\beta}=\langle \sigma^i_\alpha \sigma^j_\beta \rangle$ 
($\alpha,\beta=x,y,z$) denote two-point spin-spin functions at sites $i$ and $j$, with the expectation 
value taken over the quantum state of the system. Note that, in case of translation invariance, 
we will have that $G^k_{z}=G^{k^\prime}_{z}$ ($\forall\, k,k^\prime$) and, therefore, $b_1=b_2$. Moreover, observe 
also that the density operator given in Eq.~(\ref{rhoAB}) can be decomposed as
\begin{equation}
\rho = \frac{1}{4} \left[ I\otimes I + \sum_{i=1}^{3} \left( c_i \sigma^i \otimes \sigma^i \right) +
c_4 I \otimes \sigma^3 +  c_5 \sigma^3 \otimes I \right], 
\end{equation}
with 
\begin{eqnarray}
c_1 &=& 2z+2f, \nonumber \\
c_2 &=& 2z-2f, \nonumber \\
c_3 &=& a+d-b_1-b_2, \nonumber \\
c_4 &=& a-d-b_1+b_2,\nonumber \\
c_5 &=& a-d+b_1-b_2.
\label{gen-c}
\end{eqnarray} 
In particular, for translation invariant systems, we have that $c_4=c_5$.
In order to determine classical and quantum correlations, we first evaluate the mutual information as given by 
Eq.~(\ref{qmi1}). The eigenvalues of $\rho$ read
\begin{eqnarray}
\lambda_0 &=& \frac{1}{4} \left[ \left(1+c_3\right) + \sqrt{\left(c_4+c_5\right)^2 + \left(c_1-c_2\right)^2}\right], 
\nonumber \\
\lambda_1 &=& \frac{1}{4} \left[ \left(1+c_3\right) - \sqrt{\left(c_4+c_5\right)^2 + \left(c_1-c_2\right)^2}\right], 
\nonumber \\
\lambda_2 &=& \frac{1}{4} \left[ \left(1-c_3\right) + \sqrt{\left(c_4-c_5\right)^2 + \left(c_1+c_2\right)^2}\right], 
\nonumber \\
\lambda_3 &=& \frac{1}{4} \left[ \left(1-c_3\right) - \sqrt{\left(c_4-c_5\right)^2 + \left(c_1+c_2\right)^2}\right].
\label{qmi-ev}
\end{eqnarray}
Therefore, the mutual information is given by 
\begin{equation}
I(\rho) = S(\rho^A) + S(\rho^B) + \sum_{\alpha=0}^{3} \lambda_\alpha \log \lambda_\alpha, 
\label{qmi-final}
\end{equation}
where
\begin{eqnarray}
S(\rho^A) &=& -\left( r^A_1 \log r^A_1 + r^A_2 \log r^A_2 \right),\nonumber \\
S(\rho^B) &=& -\left( r^B_1 \log r^B_1 + r^B_2 \log r^B_2 \right),
\label{R}
\end{eqnarray}
with $r^A_1 = (1+c_5)/2$, $r^A_2 = (1-c_5)/2$, $r^B_1 = (1+c_4)/2$, and $r^B_2 = (1-c_4)/2$. 
Classical correlations can be obtained by following a procedure that is similar to those of 
Refs.~\cite{Luo:08,Dillenschneider:08}, but applying it now for the case of the general density matrix 
given by Eq.~(\ref{rhoAB}). We first introduce a set of projectors 
for a local measurement on part $B$ given by $\{ B_k = V \Pi_k V^\dagger\}$, where 
$\{\Pi_k = |k\rangle \langle k| : k=0,1\}$ is the set of projectors on the computational basis 
($|0\rangle \equiv |\uparrow\rangle$ and $|1\rangle \equiv |\downarrow\rangle$) and  
$V \in U(2)$. Note that the projectors $B_k$ represent therefore an arbitrary local measurement on $B$. 
We parametrize $V$ as 
\begin{equation}
V=\left( 
\begin{array}{cc}
\cos\frac{\theta}{2} & \sin\frac{\theta}{2} e^{-i\phi}\\ 
\sin\frac{\theta}{2} e^{i\phi} & -\cos\frac{\theta}{2} 
\end{array}
\right) ,
\label{vu2}
\end{equation}
where $0\le\theta\le \pi$ and $0\le \phi < 2\pi$. Note that $\theta$ and $\phi$ can be interpreted as the azimuthal and 
polar angles, respectively, of a qubit over the Bloch sphere.
By using Eq.~(\ref{rhok}) and the equation $\Pi_k \sigma^i \Pi_k = \delta_{i3} (-1)^k \Pi_k$, with 
$\delta_{i3}$ denoting the Kronecker symbol, we can show 
that the state of the system after measurement $\{B_k\}$ will change to one of the states
\begin{eqnarray}
\rho_0 &=& \frac{1}{2} \left( I + \sum_{j=1}^{3} q_{0j} \sigma^j \right) \otimes 
\left( V\Pi_0 V^\dagger \right), \label{rho0} \\
\rho_1 &=& \frac{1}{2} \left( I + \sum_{j=1}^{3} q_{1j} \sigma^j \right) \otimes 
\left( V\Pi_1 V^\dagger \right), 
\label{rho1}
\end{eqnarray}
where
\begin{eqnarray}
q_{k1} &=& (-1)^k \, c_1 \left[ \frac{w_1}{1+(-1)^k c_4 w_3} \right], \nonumber \\ 
q_{k2} &=& (-1)^k \, c_2 \left[ \frac{w_2}{1+(-1)^k c_4w_3} \right], \nonumber \\
q_{k3} &=& (-1)^k \, \left[ \frac{c_3 w_3 + (-1)^k c_5}{1 + (-1)^k c_4w_3}\right] , 
\end{eqnarray}
with $k=0,1$ and 
\begin{eqnarray}
w_1 &=& \sin\theta\cos\phi, \nonumber \\
w_2 &=& \sin\theta\sin\phi, \nonumber \\
w_3 &=& \cos\theta .
\end{eqnarray}
Then, by evaluating von Neumann entropy from Eqs.~(\ref{rho0}) and~(\ref{rho1}) and using that $S(V\Pi_kV^\dagger)=0$, we obtain
\begin{equation}
S(\rho_k) = - \frac{\left(1+\theta_k\right)}{2} \log \frac{\left(1+\theta_k\right)}{2} - \frac{\left(1-\theta_k\right)}{2} \log \frac{\left(1-\theta_k\right)}{2} ,
\label{vnentropy}
\end{equation}
with 
\begin{equation}
\theta_k = \sqrt{ \sum_{j=1}^{3} q_{kj}^2} .
\end{equation}
Therefore, the classical correlation for the spin pair at sites $i$ and $j$  will be given by
\begin{equation}
C(\rho) =  \max_{\{B_k\}} \left( S(\rho^A) - \frac{\left(S_0 + S_1\right)}{2}  
- c_4 w_3 \frac{\left( S_0 - S_1\right)}{2}  \right), 
\label{cc-final}
\end{equation}
where $S_k = S(\rho_k)$. 
For some cases, the maximization in Eq.~(\ref{cc-final}) can be worked out and an expression purely in terms of the 
spin correlation functions can be obtained (e.g., the XXZ and Ising chains below). In general, however, $C(\rho)$ 
has to be numerically evaluated by optimizing over the angles $\theta$ and $\phi$. 
Once classical correlation is obtained, insertion of Eqs.~(\ref{qmi-final}) and~(\ref{cc-final}) into Eq.~(\ref{qcorrel}) 
can be used to determine the quantum discord.

%%%%%%%%%%%%%%%%%%%%%%%%%%%%%%%%%%%%%%%%%%%%%%%%%%%%%%%%%%%%%%%%%%%%
\section{The XXZ spin chain} 
%%%%%%%%%%%%%%%%%%%%%%%%%%%%%%%%%%%%%%%%%%%%%%%%%%%%%%%%%%%%%%%%%%%%

\label{sec-xxz}

\begin{figure}[th]
\centering {\includegraphics[angle=0,scale=0.33]{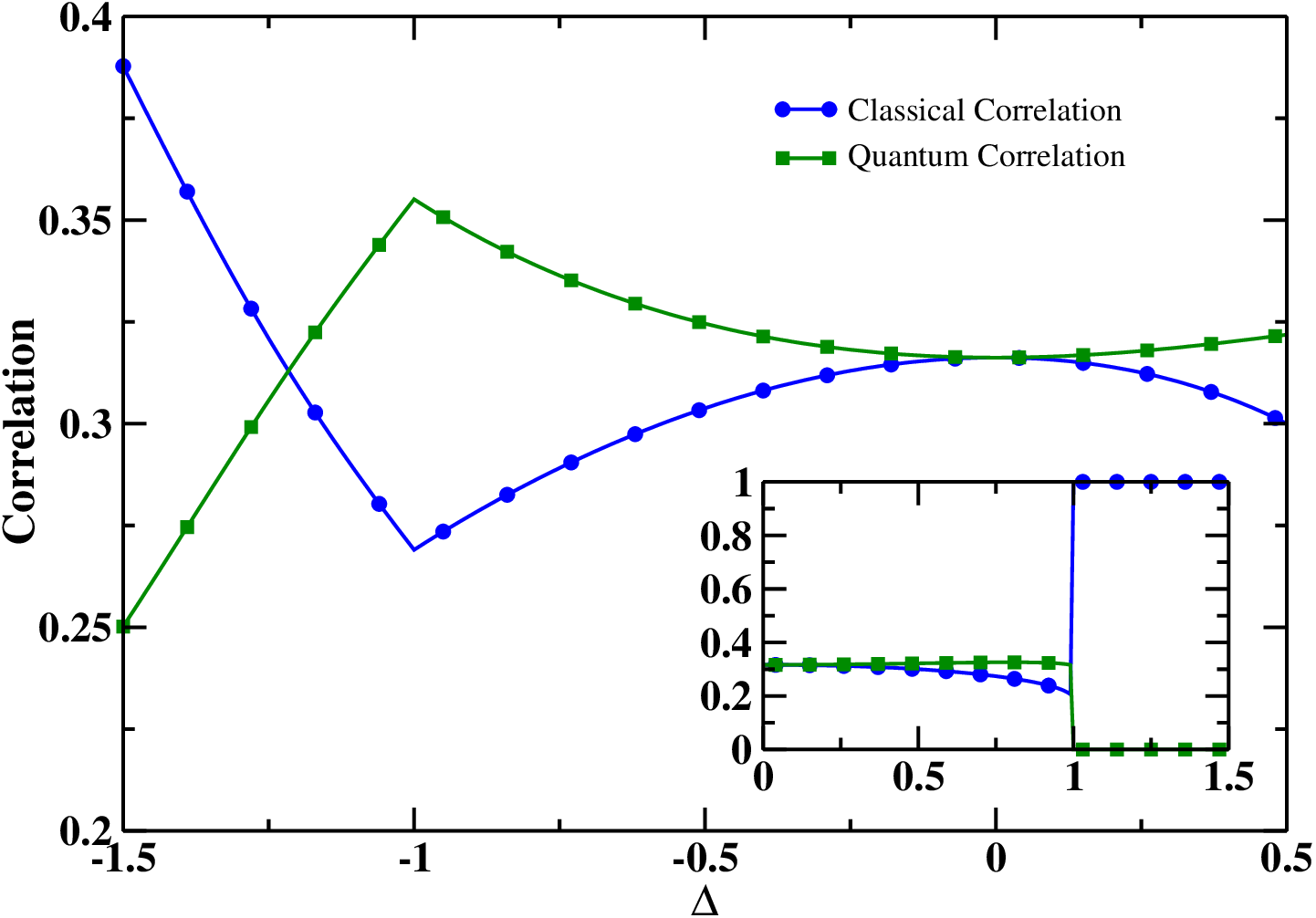}}
\caption{(Color online) Quantum and classical correlations for nearest-neighbor spins in the XXZ chain for 
$L\rightarrow \infty$. }
\label{f1}
\end{figure}

Let us illustrate the discussion of classical and quantum correlations between two spins by considering 
the XXZ spin chain, whose Hamiltonian is given by
\begin{equation}
H_{XXZ}=-\frac{J}{2} \sum_{i=1}^{L} \left( \sigma^x_i \sigma^x_{i+1} + 
\sigma^y_i \sigma^y_{i+1} + \Delta \sigma^z_i \sigma^z_{i+1} \right),
\label{HXXZ}
\end{equation}
where periodic boundary conditions are assumed, ensuring therefore translation symmetry. 
We will set the energy scale such that $J=1$ and will be interested in a nearest-neighbor spin pair at 
sites $i$ and $i+1$. Concerning 
its symmetries, the XXZ chain exhibits $U(1)$ invariance, namely, $\left[H,\sum_i \sigma_z^i\right]=0$, which provides a stronger constraint over the elements of the density matrix than the $Z_2$ symmetry. 
Indeed, $U(1)$ invariance ensures that the element $f$ 
of the reduced density matrix given by Eq.~(\ref{rhoAB}) vanishes. Moreover, the ground state has magnetization 
density $G^k_{z} = \langle \sigma_z^k \rangle=0$ ($\forall\, k$), which implies that 
\begin{eqnarray}
a &=& d = \frac{1}{4} \left(1+G_{zz}\right) \, , \nonumber \\
b_1 &=& b_2 = \frac{1}{4} \left(1-G_{zz}\right) \, , \nonumber \\
z &=& \frac{1}{4} \left(G_{xx}+G_{yy} \right), \nonumber \\
f &=& 0.
\label{relem-xxz}
\end{eqnarray}
where, due to translation invariance, we write $G_{\alpha\beta}=\langle \sigma^i_\alpha \sigma^{i+1}_\beta \rangle$ 
($\forall \, i$). 
Due to the fact that $a=d$, we will have that $c_4=c_5=0$, which considerably 
simplifies the computation of classical and quantum correlations. Moreover, we will have 
that $c_1=c_2=2z$ and $c_3 = 4a-1$. Then, the maximization procedure in Eq.~(\ref{cc-final}) can be analytically worked out~\cite{Luo:08}, yielding  
\begin{equation}
C(\rho) = \frac{\left(1-c\right)}{2} \log \left(1-c\right) + \frac{\left(1+c\right)}{2} \log \left(1+c\right),
\label{cc-xxz}
\end{equation}
with $c = \max\left(|c_1|,|c_2|,|c_3|\right)$. For the mutual information $I(\rho)$ we obtain
\begin{equation}
I(\rho) = 2 + \sum_{i=0}^{3} \lambda_i \log \lambda_i, 
\end{equation}
where
\begin{eqnarray}
\lambda_0 &=& \frac{1}{4} \left( 1-c_1-c_2-c_3 \right), \nonumber \\
\lambda_1 &=& \frac{1}{4} \left( 1-c_1+c_2+c_3 \right), \nonumber \\
\lambda_2 &=& \frac{1}{4} \left( 1+c_1-c_2+c_3 \right), \nonumber \\
\lambda_3 &=& \frac{1}{4} \left( 1+c_1+c_2-c_3 \right).
\label{qmi-ev-xxz}
\end{eqnarray}
In order to compute $C(\rho)$ and $Q(\rho)$ we write $c_1$, $c_2$, and $c_3$ in terms of the ground state 
energy density. By using the Hellmann-Feynman theorem~\cite{Hellmann:37,Feynman:39} for the XXZ  Hamiltonian~(\ref{HXXZ}), we obtain
\begin{eqnarray}
c_1 &=& c_2 = \frac{1}{2} \left(G_{xx} + G_{yy}\right) = \Delta \frac{\partial \varepsilon_{xxz}}{\partial \Delta} 
- \varepsilon_{xxz} \, , \nonumber \\
c_3 &=& G_{zz} = -2 \frac{\partial \varepsilon_{xxz}}{\partial \Delta} \, ,
\label{c-xxz}
\end{eqnarray}
where $\varepsilon_{xxz}$ is the ground state energy density 
\begin{equation}
\varepsilon_{xxz} = \frac{\langle \psi_0| H_{XXZ} |\psi_0 \rangle}{L} = - \frac{1}{2} \left(G_{xx} + 
G_{yy} + \Delta G_{zz} \right),
\label{aux-xxz}
\end{equation}
with $|\psi_0\rangle$ denoting the ground state of $H_{XXZ}$. Eqs.~(\ref{c-xxz}) and~(\ref{aux-xxz}) hold for 
a chain with an arbitrary number of sites, allowing the discussion of correlations either for finite or infinite chains. 
Indeed, ground state energy as well as its derivatives can 
be exactly determined by Bethe Ansatz technique~\cite{Yang:66}, which allows us to obtain the 
correlation functions $c_1$, $c_2$, and $c_3$. In Fig.~\ref{f1}, we plot classical and 
quantum correlations between nearest-neighbor pairs for an infinite XXZ spin chain. 

%%%%%%%%%%%%% V2: sentence slightly corrected
Note that, in the classical Ising limit $\Delta \rightarrow \infty$, we have a  
%%%%%%%%%%%%%%%%%%%%%%%%%%%%%%%%%%%%%%%%%%%%%%%
fully polarized ferromagnet. The ground state is then a doublet given by the vectors 
$|\uparrow \uparrow \cdots \uparrow\rangle$ and $|\downarrow \downarrow \cdots \downarrow\rangle$, 
yielding the mixed state 
\begin{equation}
\rho = \frac{1}{2} |\uparrow \uparrow \cdots \uparrow\rangle \langle \uparrow \uparrow \cdots \uparrow| 
+ \frac{1}{2} |\downarrow \downarrow \cdots \downarrow\rangle \langle \downarrow\downarrow \cdots \downarrow|. 
\label{xxz-bpd}
\end{equation}
Indeed, this is simply a classical probability mixing, with $C(\rho) = I(\rho) = 1$ and $Q(\rho) = 0$. 
%%%%%% V2: Sentence included
The same applies for the antiferromagnetic Ising limit $\Delta \rightarrow - \infty$, where a doubly degenerate 
ground state arises.
%%%%%%%%%%%%%%%%%%%%%%%%%%%%%%%%%%%%%%%%%%%%%%%
Moreover, observe that the classical (quantum) correlation is a minimum (maximum) 
at the infinite order QCP $\Delta=-1$. On the other hand, both correlations 
are discontinuous at the first-order QCP $\Delta=1$. This is indeed in agreement with 
the usual behavior of entanglement both at infinite and first-order QPTs. For an infinite-order QCP, 
entanglement commonly display a maximum at the QCP~\cite{Gu:03,Gu:04,Franca:06}, 
while for a first-order QCP, entanglement usually exhibits a jump at the QCP~\cite{Bose:02,Alcaraz:03}. 
Nevertheless, we note that in the specific case of the ferromagnetic QCP $\Delta=1$ and for  
pairwise entanglement measures such as concurrence~\cite{Wootters:98} and negativity~\cite{Vidal:02a},  
no jump is detected, being hidden by the operation $\max$~\cite{Yang:05}.   
It is interesting to observe the behavior of the functions $|c_1| = |c_2|$ and $|c_3|$ that govern 
the classical and quantum correlations. For $\Delta<-1$, we have that $|c_1| = |c_2| < |c_3|$, 
which means that the classical correlation is governed by $|c_3|$. For $-1<\Delta<1$, we have that  
$|c_1| = |c_2| > |c_3|$, with the crossing occurring exactly at the infinite-order QCP. Therefore, the 
correlations are governed by different parameters in different phases. For $\Delta \ge 1$, we obtain 
$|c_1| = |c_2| = 0$ and $|c_3| = 1$, which implies that $C(\rho)=1$ and $Q(\rho)=0$. These results are 
shown in Fig.~\ref{f2} below. 
\begin{figure}[th]
\centering {\includegraphics[angle=0,scale=0.33]{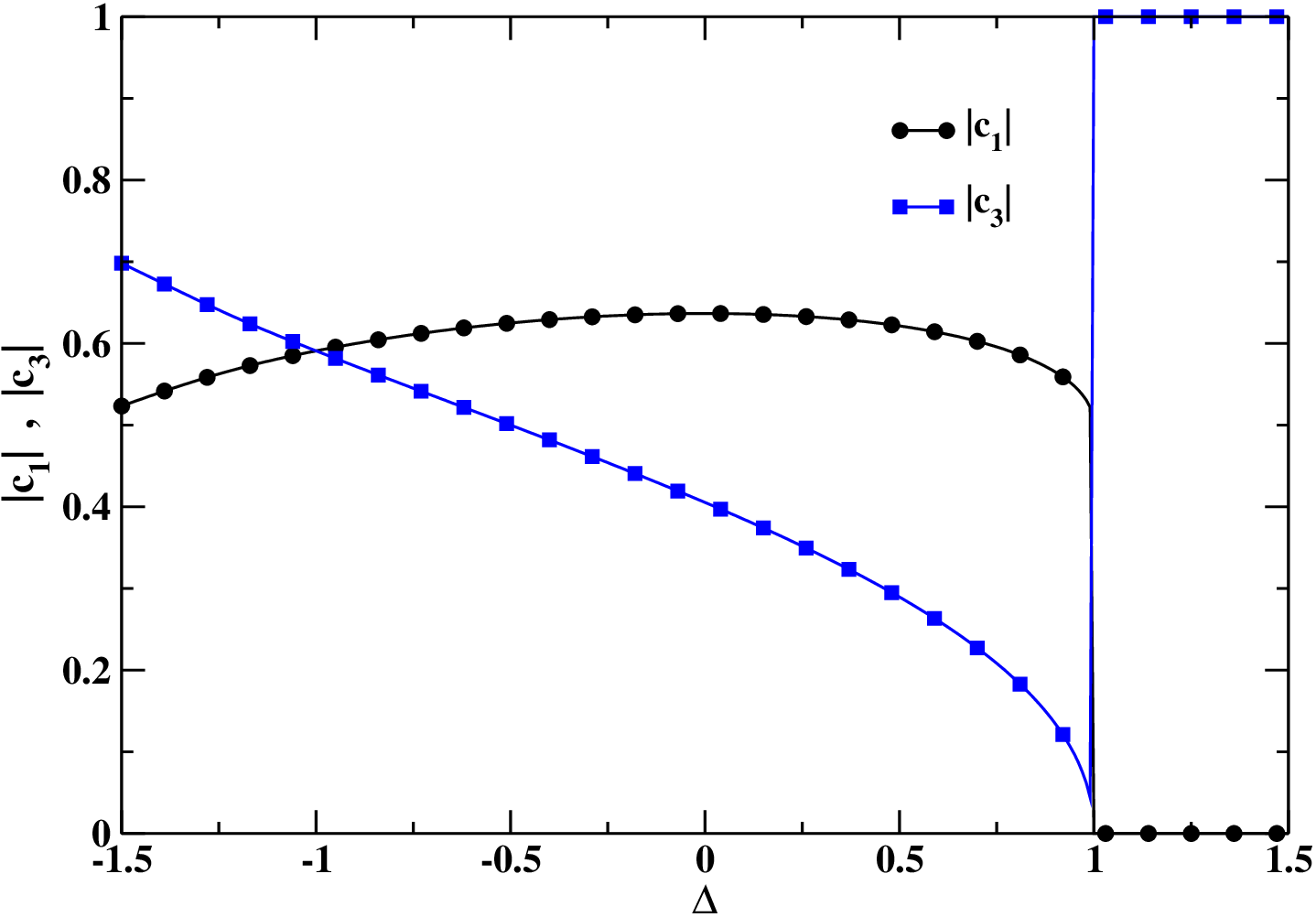}}
\caption{(Color online) Values of the parameters $|c_1| = |c_2|$ and $|c_3|$ as a function of $\Delta$. 
%%%%%%% V2: misprint corrected in the sentence below (|c_3| and |c_1| = |c_2| were uncorrectly interchanged).
The parameter $|c_3|$ governs the correlations for $\Delta<-1$ while $|c_1| = |c_2|$ is dominant in the 
%%%%%%%%%%%%%%%%%%%%%%%%%%
gapless disordered phase $-1<\Delta<1$. In the ferromagnetic phase, $|c_1| = |c_2| = 0$ and $|c_3| = 1$, 
which implies the fixed value $C(\rho)=1$ and $Q(\rho)=0$ for any $\Delta>1$.}
\label{f2}
\end{figure}

%%%%%%%%%%%%%%%%%%%%%%%%%%%%%%%%%%%%%%%%%%%%%%%%%%%%%%%%%%%%%%%%%%%%
\section{The transverse field Ising model} 
%%%%%%%%%%%%%%%%%%%%%%%%%%%%%%%%%%%%%%%%%%%%%%%%%%%%%%%%%%%%%%%%%%%%

\label{sec-ib}

\begin{figure}[th]
\centering {\includegraphics[angle=0,scale=0.33]{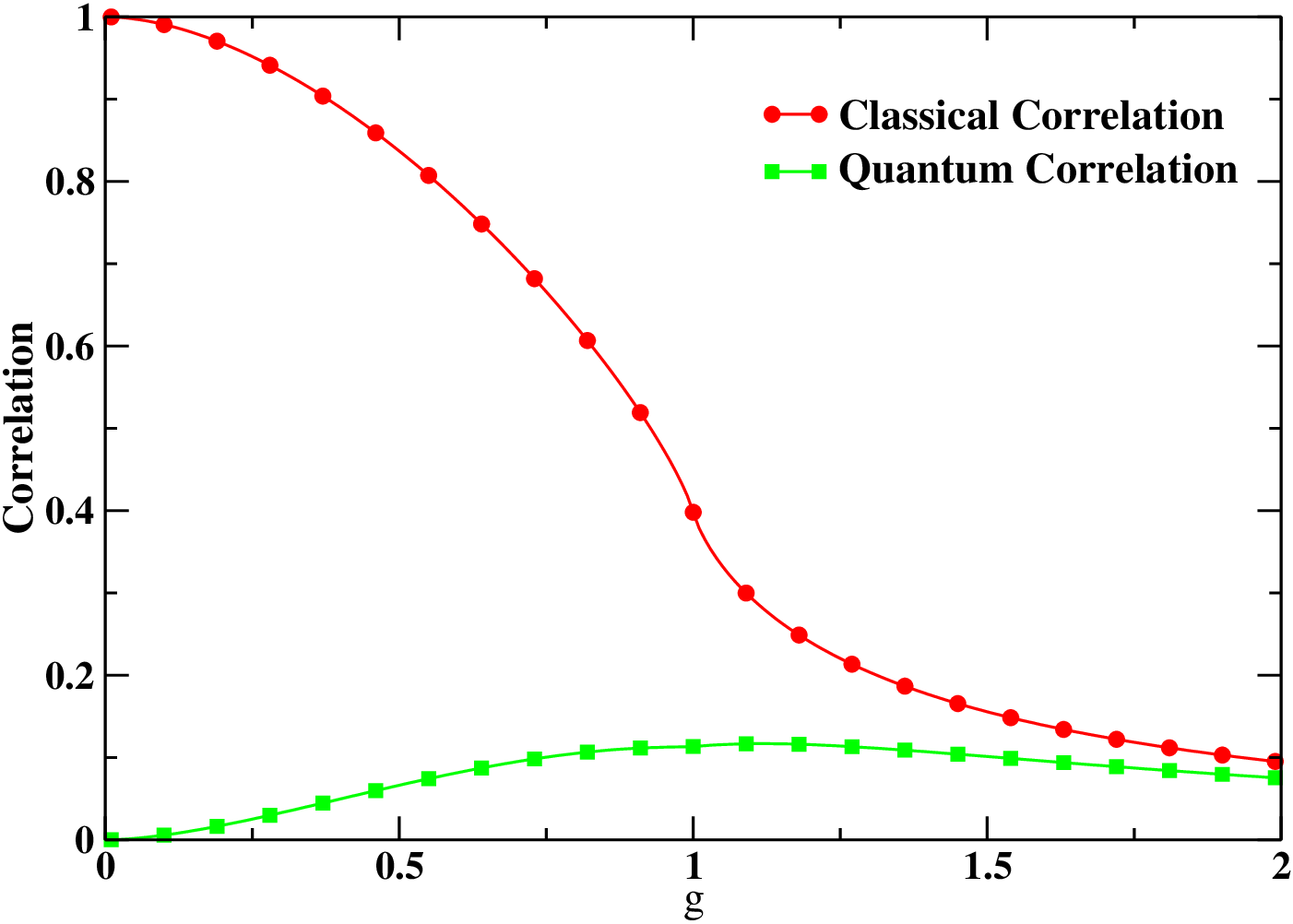}}
\caption{(Color online) Classical and quantum correlations for nearest-neighbor spins in the transverse 
field Ising model for a chain with $1024$ sites.}
\label{f3}
\end{figure}

Let us consider now the Ising chain in a transverse magnetic field, whose Hamiltonian is given by
\begin{equation}
H_{I}=- J \sum_{i=1}^{L} \left( \sigma^x_i \sigma^x_{i+1} + 
g \sigma^z_i \right),
\label{HI}
\end{equation}
with periodic boundary conditions assumed, namely, $\sigma^x_{L+1} = \sigma^x_1$. 
As before, we will set the energy scale such that $J=1$ and will be interested in a 
nearest-neighbor spin pair at sites $i$ and $i+1$.
This Hamiltonian is $Z_2$-symmetric and can be exactly diagonalized by mapping it to a spinless 
free fermion model with single orbitals. This is implemented through the Jordan-Wigner transformation 
\begin{eqnarray}
\sigma^z_i &=& 1 - 2 c^\dagger_i c_i, \nonumber \\
\sigma^x_i &=& - \prod_{j<i} \left(1-2c^\dagger_j c_j\right)\left(c_i+c_i^\dagger\right),
\label{jw}
\end{eqnarray}
where $c_i^\dagger$ and $c_i$ are the creation and annihilation fermion operators at site $i$, respectively. 
By rewriting Eq.~(\ref{HI}) in terms of $c_i^\dagger$ and $c_i$ we obtain
\begin{eqnarray}
H_{I} &=& - J \sum_{i=1}^{L} \left( c^\dagger_i c_{i+1} + c^\dagger_{i+1} c_{i} + c^\dagger_{i} c^\dagger_{i+1} + 
c_{i} c_{i+1} \right) \nonumber \\ 
&& -J g \sum_{i=1}^{L} \left(1 - 2  c^\dagger_i c_{i}  \right).
\label{HI-fermion}
\end{eqnarray}
In order to diagonalize $H_I$ we consider fermions in momentum space
\begin{eqnarray}
c_k &=& \frac{1}{\sqrt{L}} \sum_{j=1}^{L} c_j e^{-i k r_j}, \nonumber \\
c^\dagger_k &=& \frac{1}{\sqrt{L}} \sum_{j=1}^{L} c^\dagger_j e^{i k r_j}, 
\label{jw-momentum}
\end{eqnarray}
where $c^\dagger_k$ and $c_k$ are creation and annihilation fermion operators with momentum $k$, respectively, 
and $r_j$ is the fermion position at site $j$. The wave vectors $\overrightarrow{k}$ satisfy the relation 
$ka = 2\pi q / L$, where $a$ denotes the distance between two nearest-neighbor sites and 
$q = -M, -M+1, \cdots,M-1,M$, with $M = (L-1)/2$ and $L$ taken, for simplicity, as an even number. 
Then, by inverting Eq.~(\ref{jw-momentum}) and inserting the result in Eq.~(\ref{HI-fermion}), we obtain
\begin{eqnarray}
H_{I} &=& J \sum_{k} \left[ 2 \left( g - \cos ka \right) c^\dagger_k c_{k} \right. \nonumber \\
&& \left. + i \sin ka \left( c^\dagger_{-k} c^\dagger_{k} + c_{-k} c_{k} \right) - g \right].
\label{HI-momentum}
\end{eqnarray}
Diagonalization is then obtained by eliminating the terms 
$c^\dagger_{-k} c^\dagger_{k}$ and  $c_{-k} c_{k}$ from the Hamiltonian given by Eq.~(\ref{HI-momentum}), which 
do not conserve the particle number. This is indeed achieved through   
the Bogoliubov transformation in which new fermion operators $\gamma_k$ 
and $\gamma^\dagger_k$ are introduced as linear combination of $c_k$ and $c^\dagger_k$ 
\begin{eqnarray}
\gamma_k &=& u_k c_k - i v_k c^\dagger_{-k}, \nonumber \\
\gamma^\dagger_k &=& u_k c^\dagger_k + i v_k c_{-k},
\label{BT}
\end{eqnarray}
where $u_k$ and $v_k$ are real numbers parametrized by 
$u_k = \sin \frac{\theta_k}{2}$ and $v_k = \cos \frac{\theta_k}{2}$. 
This parametrization naturally arises as a consequence of the fermionic algebra 
$\{ \gamma_k , \gamma^\dagger_{k^\prime} \} = \delta_{k k^\prime}$, 
$\{ \gamma^\dagger_k , \gamma^\dagger_{k^\prime} \} = \{ \gamma_k , \gamma_{k^\prime} \} = 0$, with 
$\delta_{k k^\prime}$ standing for the Kronecker delta symbol. Moreover, to recast the Hamiltonian in a 
diagonal form we define $\theta_k$ by demanding that $\tan \theta_k = \sin ka / (g - \cos ka)$. 
Therefore, by expressing $H_I$ in terms of Bogoliubov fermions and by imposing the trace invariance 
of the Hamiltonian, Eq.~(\ref{HI-momentum}) becomes
\begin{equation}
H_{I} = \sum_k \varepsilon_k \left( \gamma^\dagger_k \gamma_k - \frac{1}{2} \right) , 
\label{HI-B}
\end{equation}
with  $\varepsilon_k = 2J \sqrt{1+g^2-2g \cos ka}$. Hamiltonian (\ref{HI-B}) is diagonal, with ground state 
given by the $\gamma$-fermion vacuum. The procedure above also applies for the evaluation of the 
matrix elements of the reduced density operator given by Eq.~(\ref{relem}), which amounts for the computation 
of the magnetization density $G_z$ and the two-point functions $G_{\alpha \beta}$. 
This can be achieved by using that $G_{zz} = G_z^2 - G_{xx}G_{yy}$~\cite{Pfeuty:70} and by 
expressing the remaining correlation functions as 
\begin{eqnarray}
G_{xx} &=& \frac{2}{L} \sum_{q=-M}^{M} \left[ \cos\left(\frac{2\pi q}{L}\right) v_q^2 +  
\sin\left(\frac{2\pi q}{L}\right) u_q v_q \right], \nonumber \\
G_{yy} &=& \frac{2}{L} \sum_{q=-M}^{M} \left[ \cos\left(\frac{2\pi q}{L}\right) v_q^2 -  
\sin\left(\frac{2\pi q}{L}\right) u_q v_q \right], \nonumber \\ 
G_z &=& \frac{1}{L} \sum_{q=-M}^{M} \left(1 - 2 v_q^2\right),
\label{g-gamma}
\end{eqnarray}
where
\begin{eqnarray}
u_q v_q &=& \frac{1}{2} \frac{\sin\left(\frac{2\pi q}{L}\right)}{\sqrt{1+g^2-2g\cos\left(\frac{2\pi q}{L}\right)}} 
\nonumber \\
v_q^2 &=& \frac{1}{2} \left[1 - \frac{\left(g - \cos\left(\frac{2\pi q}{L}\right)\right)}{\sqrt{1+g^2-2g\cos\left(\frac{2\pi q}{L}\right)}}\right].
\label{uv-final}
\end{eqnarray} 
Hence, we exactly determine the two-spin reduced density matrix. 
Classical and quantum correlations can then be directly obtained from Eqs.~(\ref{ccorrel}) and 
(\ref{qcorrel}). By numerically computing the classical correlation in Eq.~(\ref{cc-final}) for nearest-neighbor spin 
pairs at sites $i$ and $i+1$, we can show that the maximization is achieved for any $g$ by the choice $\theta=\pi/2$ 
and $\phi=0$. Then, the measurement that maximizes $J(\rho:\{B_k\})$ is given by $\{|+\rangle\langle+|, |-\rangle\langle-|\}$, 
with $|+\rangle$ and $|-\rangle$ denoting the up and down spins in the $x$ direction, namely, $|\pm\rangle = (|\uparrow\rangle \pm |\uparrow\rangle)/\sqrt{2}$. This numerical observation implies that $w_1=1$, $w_2=w_3=0$. Therefore, Eq.~(\ref{cc-final}) is ruled by the spin 
functions $c_1 = G^{i,i+1}_{xx}$ and $c_4 = c_5 = G^i_z$, i.e. 
\begin{equation}
C(\rho) =  H_{bin}\left(p_1\right) -  H_{bin}\left(p_2\right)
\label{cc-final-ib}
\end{equation}
where $H_{bin}$ is the binary entropy
\begin{equation}
H_{bin} (p) = - p \log p - \left(1-p\right) \log \left(1-p\right)
\label{H-bin}
\end{equation}
and  
\begin{eqnarray}
p_1 &=& \frac{1}{2}\left(1+G^i_z\right), \nonumber \\
p_2 &=& \frac{1}{2}\left(1+\sqrt{\left(G^{i,i+1}_{xx}\right)^2 + \left(G^{i}_{z}\right)^2}\right)
\label{p-ib}
\end{eqnarray}
We plot $C(\rho)$ and $Q(\rho)$ in Fig.~\ref{f3} for a chain with $1024$ sites. 
Note that, for $g=0$ the system is a classical Ising chain, whose 
ground state is a doublet given by the vectors $|++ \cdots +\rangle$ and $|-- \cdots -\rangle$. 
Therefore, the system is in the 
mixed state 
\begin{equation}
\rho = \frac{1}{2} |++ \cdots +\rangle \langle ++ \cdots +| + \frac{1}{2} 
|-- \cdots -\rangle \langle -- \cdots -|, 
\label{ib-bpd}
\end{equation}
with $C(\rho) = I(\rho) = 1$ and $Q(\rho) = 0$. 
On the other hand, in the limit $g \rightarrow \infty$ the system is a paramagnet 
(vanishing magnetization in the $x$ direction), with all spins in state $|\uparrow\rangle$. 
Therefore the system will be described by the density operator 
\begin{equation}
\rho = |\uparrow \uparrow \cdots \uparrow \rangle\langle \uparrow\uparrow\cdots \uparrow|,
\end{equation}
which is a pure separable state, containing neither classical nor quantum correlations. 

\begin{figure}[th]
\centering {\includegraphics[angle=0,scale=0.33]{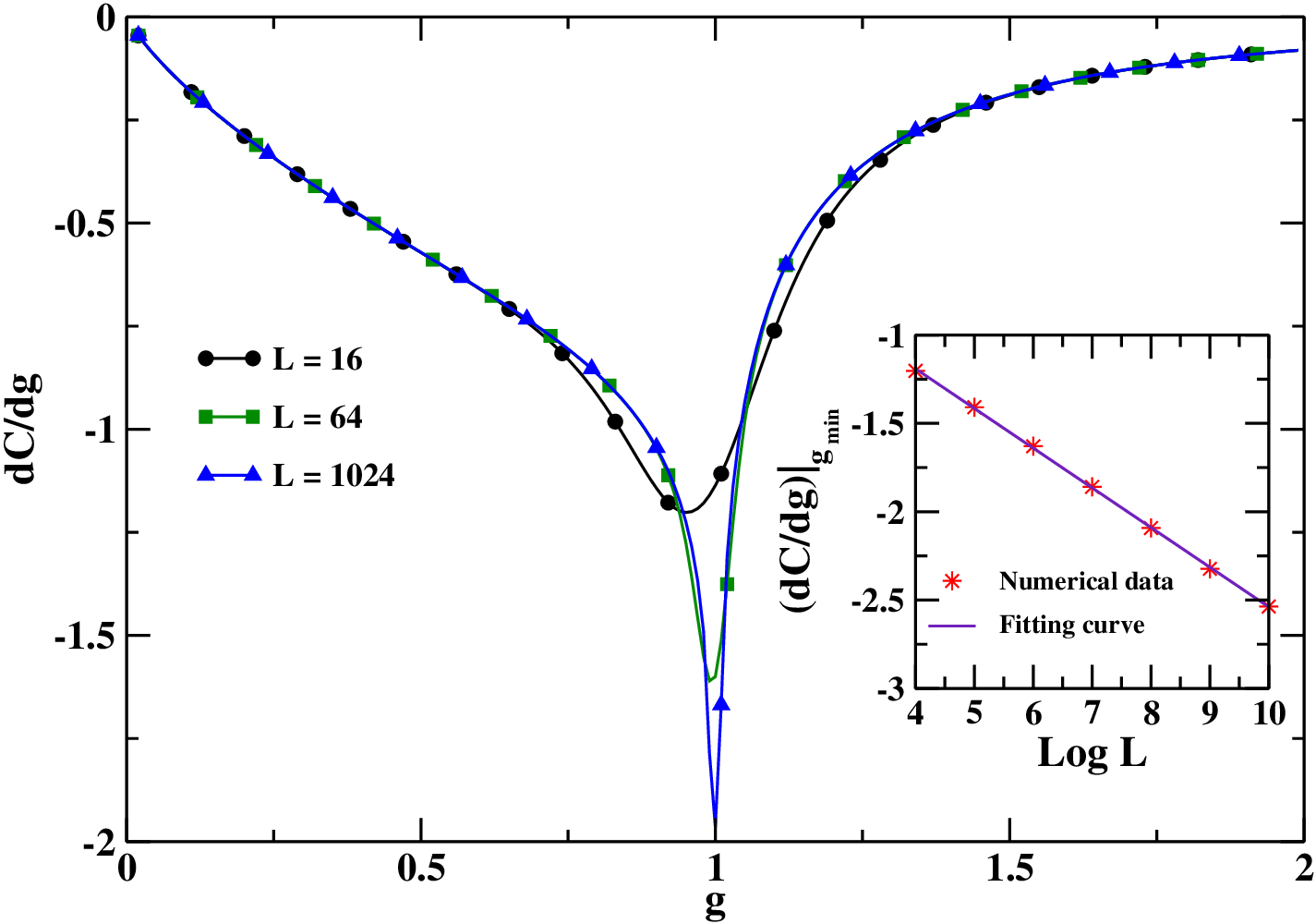}}
\caption{(Color online) First derivative of the classical correlation for nearest-neighbor 
spins with respect to $g$  in the transverse field Ising chain for different lattice sizes $L$. 
The derivative of $C$ has a pronounced minimum at $g_{min}$, which tends to the QCP $g=1$ as 
$L\rightarrow \infty$. Inset: $dC/dg$ taken at $g_{min}$ exhibits a logarithmic divergence 
fitted by $\left.(dC/dg)\right|_{g_{min}} = -0.29161 - 0.22471 \log L$.}
\label{f4}
\end{figure}

\begin{figure}[th]
\centering {\includegraphics[angle=0,scale=0.34]{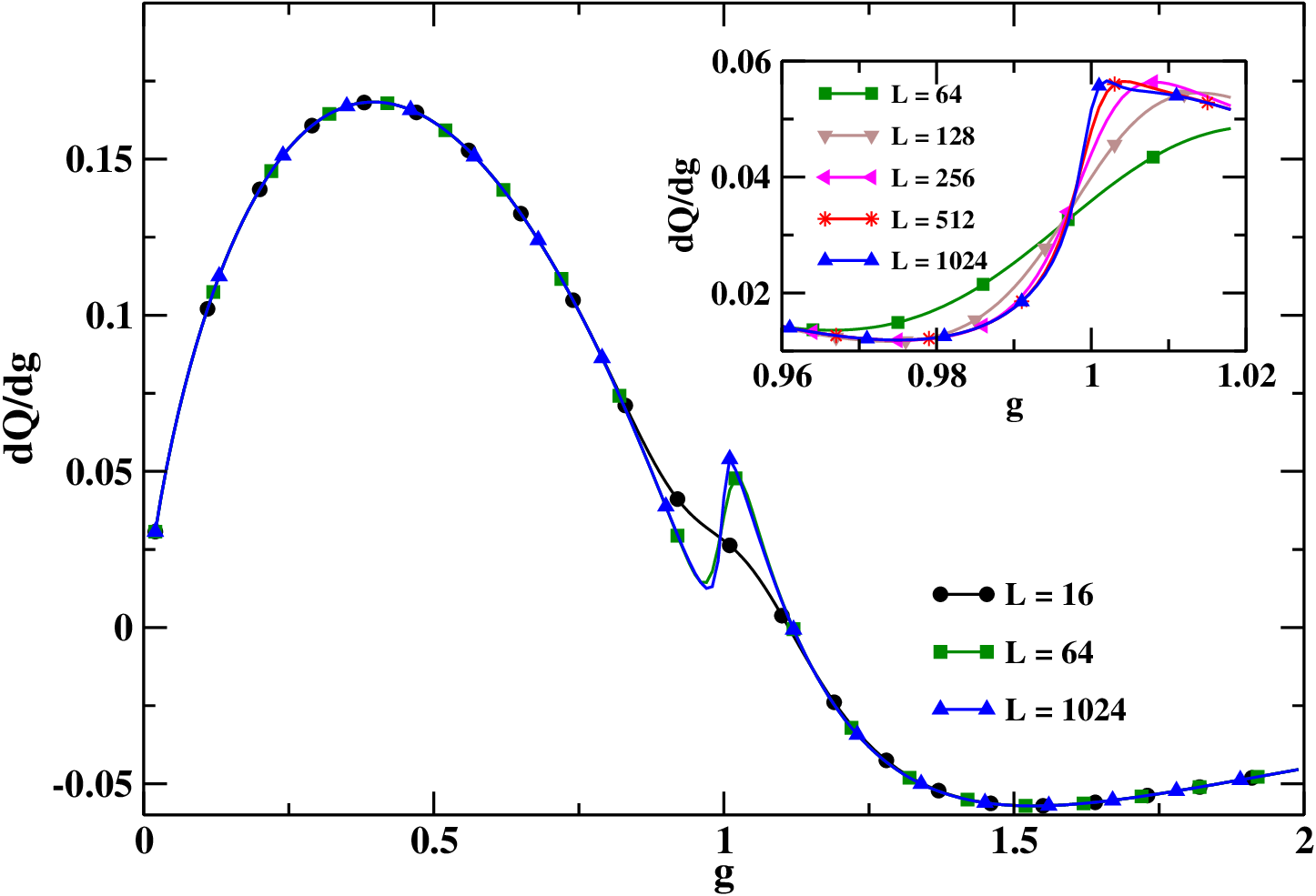}}
\caption{(Color online) First derivative of the quantum correlation for nearest-neighbor 
spins with respect to $g$ in the transverse field Ising chain for different lattice sizes $L$. 
Inset: $dQ/dg$ presents an inflexion point that tends to the QCP $g=1$ as $L\rightarrow \infty$.}
\label{f5}
\end{figure}

\begin{figure}[th]
\centering {\includegraphics[angle=0,scale=0.33]{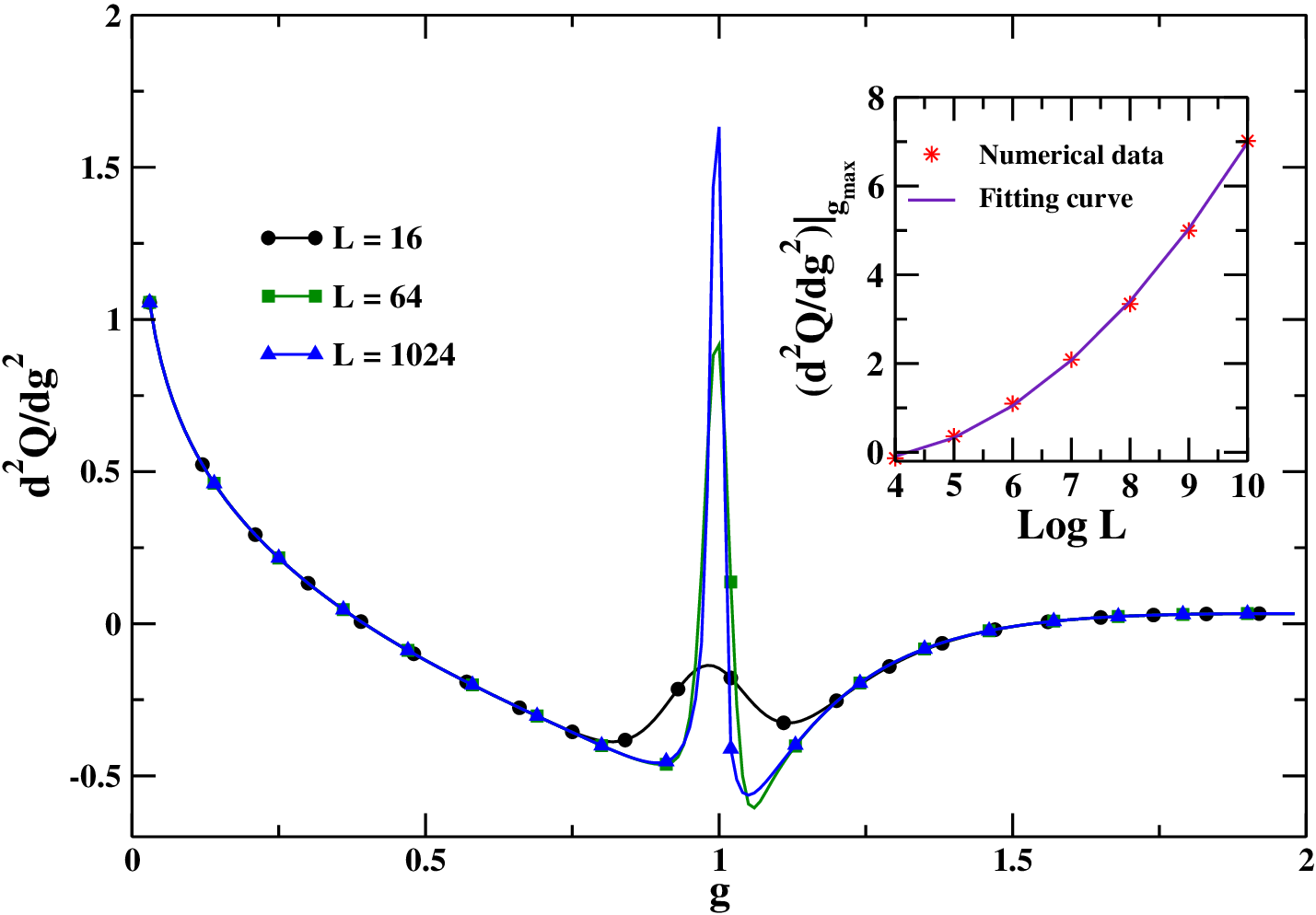}}
\caption{(Color online) Second derivative of the quantum correlations for nearest-neighbor spins 
with respect to $g$ in the transverse field Ising chain for different lattice sizes $L$. 
Observe that the second derivative of $Q$ has a pronounced maximum 
at $g_{max}$, which tends to the QCP $g=1$ as $L\rightarrow \infty$. Inset: 
$d^2 C/dg^2$ taken at $g_{max}$ exhibits a quadratic logarithmic divergence fitted by 
$\left.(d^2Q/dg^2)\right|_{g_{max}} = 1.268 - 0.94712 \log L + 0.15176 \log^2 L$.}
\label{f6}
\end{figure}

The QPT from ferromagnetic to paramagnetic state is a second-order QPT and occurs at $g=1$. 
Signatures of this QPT can be found out by looking at the derivatives of either classical or 
quantum correlations. Indeed, the QPT can be identified as a pronounced minimum of the first 
derivative of the classical correlation, which is exhibited in Fig.~\ref{f4}. Note that the minimum 
logarithmically diverges at $g=1$ as the thermodynamic limit is approached (see inset of Fig.~\ref{f4}). 
In the case of quantum correlations, its first derivative shows an inflexion point around $g=1$, as displayed in 
Fig.~\ref{f5}. Indeed, by looking at its second derivative in Fig.~\ref{f6}, 
the QPT is identified by a pronounced maximum, which shows quadratic logarithmic divergence at $g=1$ 
as the thermodynamic limit is approached (see inset of Fig.~\ref{f6}). 

\begin{figure}[th]
\centering {\includegraphics[angle=0,scale=0.35]{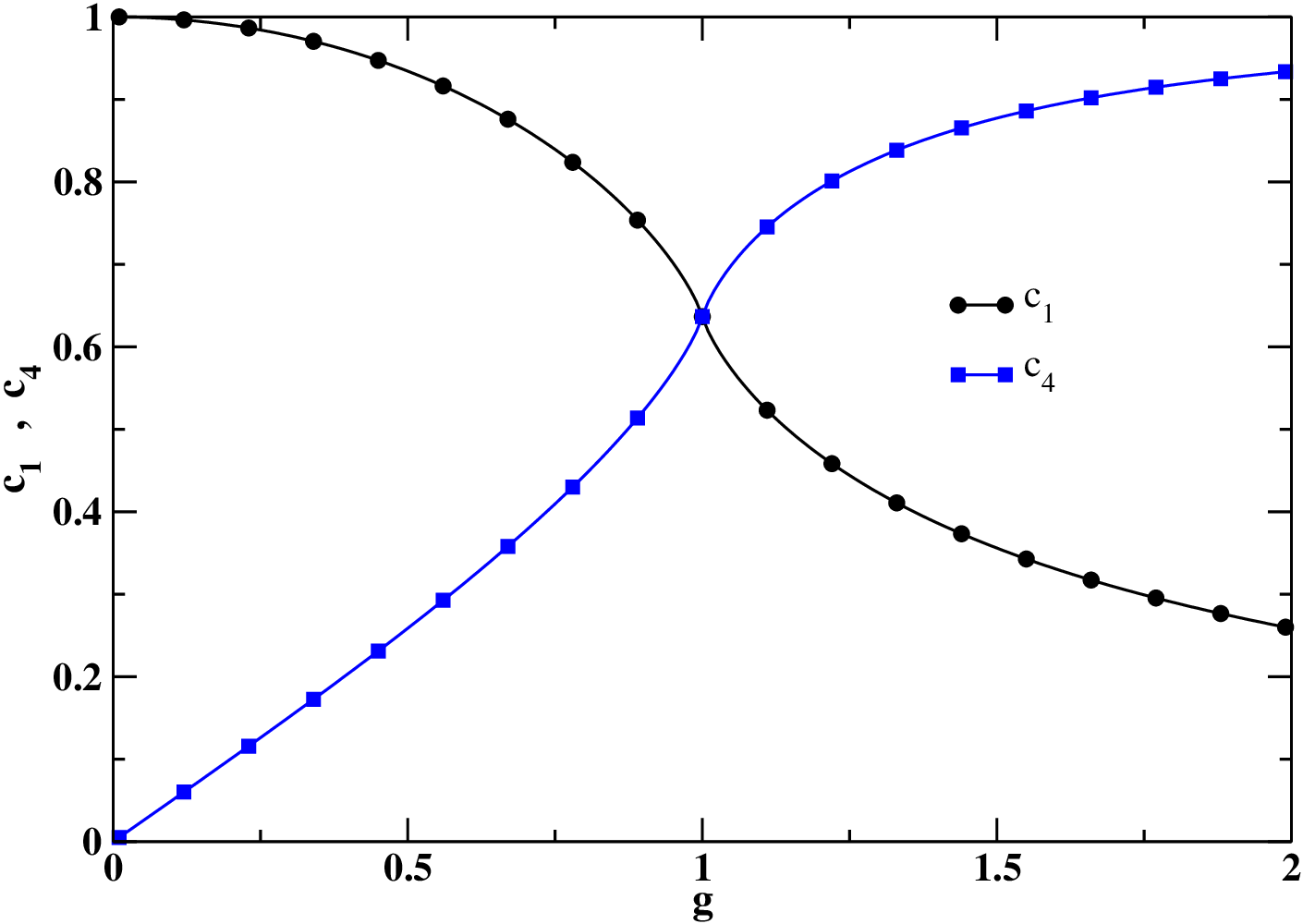}}
\caption{(Color online) Spin functions $c_1 = G^{i,i+1}_{xx}$ and $c_4 = G^i_z$ as a function of $g$. 
These functions govern the behavior of correlations, exhibiting a crossing exactly at the QCP. The 
plot is for a chain with $L=1024$ sites.}
\label{f7}
\end{figure}

The behavior of the quantum discord is therefore rather different from the entanglement behavior, 
whose first derivative is already divergent at the QCP. 
%%%%% V2: Sentence rewritten
Remarkably, the scaling of pairwise entanglement derivative in this case 
(see e.g. Refs.~\cite{Osterloh:02,Wu:04}) is much closer to the scaling of 
the classical correlation derivative (as given by Fig.~\ref{f4}) than that of the quantum 
correlation derivative (as given by Fig.~\ref{f5}). 
%%%%%%%%%%%%%%%%%%%%%%%%%%%
As in the case of the XXZ model, 
it is interesting to observe that the spin 
functions $c_1 = G^{i,i+1}_{xx}$ and $c_4 = c_5 = G^i_z$, which govern the correlations in the Ising chain 
[see Eqs.~(\ref{cc-final-ib})-(\ref{p-ib})], exhibit a crossing at the QCP. This is shown in Fig.~\ref{f7} for 
a chain with $1024$ sites.

%%%%%%%%%%%%%%%%%%%%%%%%%%%%%%%%%%%%%%%%%%%%%%%%%%%%%%%%%%%%%%%%%%%%
\section{The LMG model} 
%%%%%%%%%%%%%%%%%%%%%%%%%%%%%%%%%%%%%%%%%%%%%%%%%%%%%%%%%%%%%%%%%%%%

\label{sec-lip}

\begin{figure}[th]
\centering {\includegraphics[angle=0,scale=0.35]{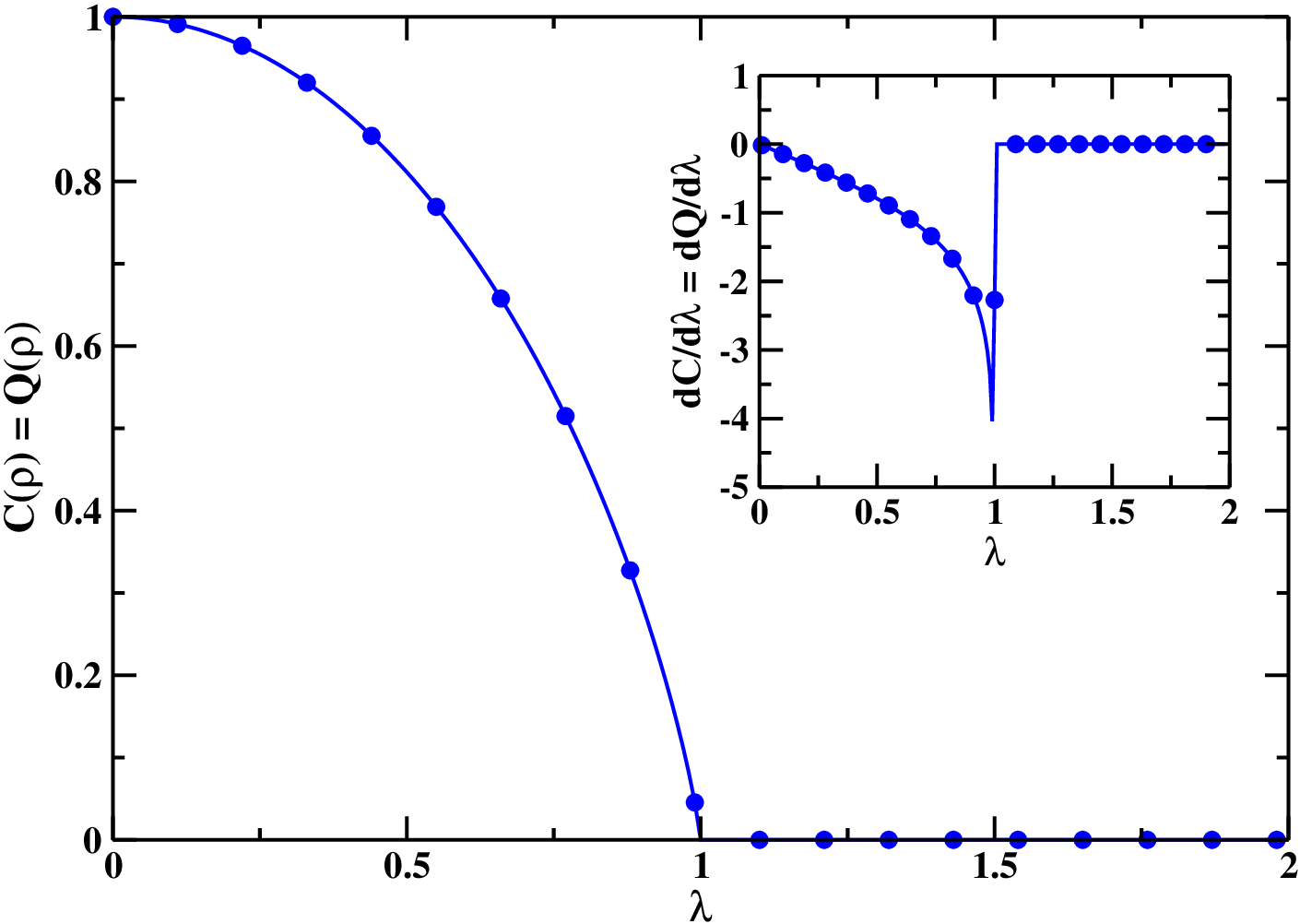}}
\caption{(Color online) Classical and quantum correlations between modes $(+m)$ and $(-m)$ in the HF ground state of 
the Lipkin model. Inset: Derivative of correlations is nonanalytic at $\lambda=1$. }
\label{f8}
\end{figure}

The discussion of correlations above can also be applied in collective systems. 
As an illustration, we will consider here the LMG model~\cite{LMG:65}, which 
describes a two-level Fermi system $\{|+\rangle ,|-\rangle \}$, with each 
level having degeneracy $\Omega$. The Hamiltonian for LMG model is given by
\begin{eqnarray}
&&H = \lambda \sum_{m=1}^{\Omega} \frac{1}{2} \left(c_{+m}^{\dagger } c_{+m} - c_{-m}^{\dagger } c_{-m}\right) \nonumber \\
&&-\frac{1}{2N} \sum_{m,n=1}^{\Omega} \left( c_{+m}^{\dagger} c_{-m} c_{+n}^{\dagger} c_{-n} + 
c_{-n}^{\dagger} c_{+n} c_{-m}^{\dagger} c_{+m} \right).\nonumber \\
\end{eqnarray}
The operators $c_{+m}^{\dagger }$ and $c_{-m}^{\dagger }$ create a particle in the upper and lower levels,
respectively. This Hamiltonian can be taken as describing an effective model for many-body systems, with one 
level just below the Fermi level and and the other level just above, with the level below being filled with 
$\Omega$ particles~\cite{Ring:book}. Alternatively, the LMG model can be seen as a one-dimensional 
ring of spin-1/2 particles with infinite range interaction between pairs. 
Indeed, the Hamiltonian can be rewritten as 
\begin{equation}
H=\lambda S_{z}-\frac{1}{N}\left(S_{x}^{2}-S_{y}^{2}\right), 
\end{equation}
where $S_{z}=\sum_{m=1}^{N}\frac{1}{2}(c_{+m}^{\dagger }c_{+m}-c_{-m}^{\dagger }c_{-m})$ and 
$S_{x}+iS_{y}=\sum_{m=1}^{N}c_{+m}^{\dagger }c_{-m}$~\cite{Ring:book}. 
%%%%%%%%% V2: Sentence rewritten
The system undergoes a second-order QPT at $\lambda=1$. 
As $\Omega \rightarrow \infty$, the ground state, as given by the Hartree-Fock (HF) approach, reads
%%%%%%%%%%%%%%%%%%%%%%%%%%%%%%%%
\begin{equation}
|HF\rangle = \prod_{m=1}^{\omega} a^{\dagger}_{0m} |-\rangle, 
\label{HF-gs}
\end{equation}
where we have introduced new levels labelled by $0$ and $1$ governed by the operators
\begin{eqnarray}
a^{\dagger}_{0m} &=& \cos\alpha \,c_{-m}^{\dagger} + \sin\alpha \,c_{+m}^{\dagger}, \nonumber \\
a^{\dagger}_{1m} &=& - \sin\alpha\, c_{-m}^{\dagger} + \cos\alpha \,c_{+m}^{\dagger}.
\label{nl-lipkin}
\end{eqnarray}
In Eq.~(\ref{nl-lipkin}), $\alpha$ is a variational parameter to be adjusted in order to minimize energy, which 
is achieved according to the choice
\begin{eqnarray}
\lambda < 1 &\Rightarrow& \cos 2\alpha = \lambda, \nonumber \\
\lambda \ge 1 &\Rightarrow& \alpha = 0.
\end{eqnarray}
%%%%%%%%%% V2: New sentence and 3 refs included.
Despite being an approximation, the HF ground state provides the exact description of the critical point 
(for recent discussions of the exact spectrum of the LMG model, see Refs.~\cite{Ribeiro:07,Ribeiro:08}).
%%%%%%%%%%%%%%%%%%%%%%%%%%%%%%%%%%%%%%%%%%%%%%%%
The pairwise density operator for general modes $i\equiv (+m)$ and $j\equiv (-n)$ is given by 
\begin{equation}
\rho_{i,j}=\left( 
\begin{array}{cccc}
\langle M_{i} M_{j} \rangle & 0 & 0 & 0 \\ 
0 & \langle M_{i} N_{j} \rangle & \langle c^\dagger_{i} c_{j} \rangle & 0 \\ 
0 & \langle c^\dagger_{j} c_{i} \rangle & \langle N_{i} M_{j} \rangle & 0 \\ 
0 & 0 & 0 & \langle N_{i} N_{j} \rangle
\end{array}
\right) ,   \label{rhoAB-lipkin}
\end{equation}
where $M_k = 1 - N_k$ and $N_k = c^\dagger_k c_k$, with $k = i,j$. By evaluating the 
matrix elements of $\rho$ for the HF ground state, we obtain
\begin{eqnarray}
\langle M_{+m} M_{-n} \rangle &=& \sin^2\alpha \cos^2\alpha \left(1-\delta_{mn}\right), \nonumber \\
\langle M_{+m} N_{-n} \rangle &=& \cos^2\alpha \delta_{mn} + \cos^4\alpha \left( 1 - \delta_{mn} \right) \nonumber \\
\langle N_{+m} M_{-n} \rangle &=& \sin^2\alpha \delta_{mn} + \sin^4\alpha \left( 1 - \delta_{mn} \right) \nonumber \\
\langle N_{+m} N_{-n} \rangle &=& \sin^2\alpha \cos^2\alpha \left( 1 - \delta_{mn} \right) \nonumber \\
\langle c^\dagger_{+m} c_{-n} \rangle &=& \sin\alpha \cos\alpha \delta_{mn} \nonumber \\
\langle c^\dagger_{-n} c_{+m} \rangle &=& \sin\alpha \cos\alpha \delta_{mn} .
\label{cf-lipkin}
\end{eqnarray}
Note that Eq.~(\ref{rhoAB-lipkin}) displays $Z_2$ symmetry and, therefore, classical and 
quantum correlations can be computed by using Eq.~(\ref{cc-final}). Note also that, for 
$m\ne n$, the density matrix is diagonal and the state is completely pairwise uncorrelated. 
On the other hand, for $m=n$, there is an equal amount of classical and quantum correlations 
between the modes. These correlations vanish for $\lambda>1$, which is the fully polarized state. 
The result is plotted in Fig.~\ref{f8}. We can then observe that the derivatives of both 
classical correlation and quantum discord exhibit a signature of the QPT (see inset of Fig.~\ref{f8}). 
These signatures are in agreement with the caracterizations in terms of entanglement~\cite{Vidal:04,Wu:06} and Fisher information~\cite{Ma:09}.

%%%%%%%%%%%%%%%%%%%%%%%%%%%%%%%%%%%%%%%%%%%%%%%%%%%%%%%%%%%%%%%%%%%%
\section{Conclusion} 
%%%%%%%%%%%%%%%%%%%%%%%%%%%%%%%%%%%%%%%%%%%%%%%%%%%%%%%%%%%%%%%%%%%%

\label{sec-con}

%%%%%%%%%%% V2: Conclusion rewritten to address Referee's suggestions
In conclusion, we have investigated the behavior of pairwise correlations in general 
$Z_2$ symmetric systems, splitting up their classical and quantum contributions. This 
allowed for the treatment of spin systems such as the XXZ model and the transverse 
field Ising chain, where the maximization required for the evaluation of the correlations 
has been analytically worked out. Moreover, we have identified the spin functions that 
govern the correlations and also discussed their behavior for finite size chains. As a 
further application, we have used our approach to investigate the case of a 
collective model, given by the LMG Hamiltonian.

As it was shown, both classical correlation and quantum discord display signatures of the critical 
behavior of the system for the cases of first-order, second-order, and infinite-order QPTs. For 
first-order QPTs, as illustrated by the ferromagnetic point of the XXZ spin chain, we found that 
both classical correlation and quantum discord display a jump at the critical point, which closely 
resembles the behavior of entanglement. For second-order QPTs, although both kind of correlations 
exhibit signatures of the QPTs, the derivatives of the quantum discord show a scaling that is rather 
different from the entanglement behavior in the transverse field Ising model (see, e.g., 
Refs.~\cite{Osterloh:02,Wu:04}). It is remarkable that, for this model, the first derivative of pairwise 
entanglement with respect to the parameter that drives the QPT exhibits a scaling that is much 
closer to the scaling of the first derivative of the {\it classical} correlation. 
For infinite-order QPTs, as given by the antiferromagnetic point of the XXZ spin chain, 
classical correlation is a minimum at the QCP while quantum discord is a maximum. 
Whether or not this might be a general feature of infinite-order QPTs is still under analysis. 

Further investigations including correlations between blocks of particles and the effect of temperature 
may be interesting to establish a precise comparison between (classical and quantum) correlations 
and entanglement at QPTs. Moreover, dynamics in open quantum systems~\cite{Shabani:09,Werlang:09,Maziero:09} 
may also provide an interesting scenario for the discussion of the properties of the correlations and 
its implications for phase transitions. Such topics are left for a future research. 
%%%%%%%%%%%%%%%%%%%%%%%%%%%%%%%%%%%%%%%%%%%%%%%%%%%%%%%%%%%%%%%%%%%%%%%%

%%%%%%%%%%%%%%%%%%%%%%%%%%%%%%%%%%%%%%%%%%%%%%%%%%%%%%%%%%%%%%%%%%%%
\subsection*{Acknowledgments}
%%%%%%%%%%%%%%%%%%%%%%%%%%%%%%%%%%%%%%%%%%%%%%%%%%%%%%%%%%%%%%%%%%%%

This work was supported by the Brazilian agencies MCT/CNPq and FAPERJ.


\begin{thebibliography}{9}

\bibitem{Sachdev:book}
{S. Sachdev}, {\em {Quantum Phase Transitions}}, {Cambridge University Press},
{Cambridge, U.K.}, 2001.

\bibitem{Continentino:book}
{M. A. Continentino}, {\em {Quantum Scaling in Many-Body Systems}}, World Scientific, Singapore, 2001.

\bibitem{Amico:08} L. Amico, R. Fazio, A. Osterloh, and V. Vedral, Rev. Mod. Phys. {\bf 80}, 517 (2008).

\bibitem{Osterloh:02} A. Osterloh, L. Amico, G. Falci, and R. Fazio, Nature {\bf 416}, 608 (2002).

\bibitem{Wu:04} L.-A. Wu, M. S. Sarandy, and D. A. Lidar, Phys. Rev. Lett. {\bf 93}, 250404 (2004).

\bibitem{Vidal:03} G. Vidal, J. I. Latorre, E. Rico, and A. Kitaev, Phys. Rev. Lett. {\bf 90}, 227902 (2003).

\bibitem{Korepin:04} V. E. Korepin, Phys. Rev. Lett. {\bf 92}, 096402 (2004).

\bibitem{Calabrese:04} P. Calabrese and J. Cardy, J. Stat. Mech. {\bf 0406}, 002 (2004).

\bibitem{Refael:04} G. Refael and J. E. Moore, Phys. Rev. Lett. {\bf 93}, 260602 (2004).

\bibitem{Saguia:07} A. Saguia, M. S. Sarandy, B. Boechat, and M. A. Continentino, 
Phys. Rev. A {\bf 75}, 052329 (2007).

\bibitem{Lin:07} Y.-C. Lin, F. Igl\'oi, and H. Rieger, Phys. Rev. Lett. {\bf 99}, 147202 (2007).

\bibitem{Hur:07} K. Le Hur, P. Doucet-Beaupr\'e, and W. Hofstetter, 
Phys. Rev. Lett. {\bf 99}, 126801 (2007).

\bibitem{Plenio:05} M. B. Plenio, J. Eisert, J. Drei\ss ig, and M. Cramer, 
Phys. Rev. Lett. {\bf 94}, 060503 (2005).

\bibitem{Wolf:06} M. M. Wolf, Phys. Rev. Lett. {\bf 96}, 010404 (2006).

\bibitem{Yu:08} R. Yu, H. Saleur, and S. Haas, Phys. Rev. B {\bf 77}, 140402(R) (2008).

\bibitem{Wolf:08} M. M. Wolf, F. Verstraete, M. B. Hastings, and J. I. Cirac, 
Phys. Rev. Lett. {\bf 100}, 070502 (2008).

\bibitem{Ollivier:02} H. Ollivier and W. Zurek, Phys. Rev. Lett. {\bf 88}, 017901 (2002).

\bibitem{Luo:08} S. Luo, Phys. Rev. A {\bf 77}, 042303 (2008).

\bibitem{Kaszlikowski:08} D. Kaszlikowski, A. Sen(De), U. Sen, V. Vedral, and A. Winter, 
Phys. Rev. Lett. {\bf 101}, 070502 (2008).

%%%%%%%%%%% V2: Ref. included
\bibitem{Horodecki:08} A. Grudka, M. Horodecki, P. Horodecki, and R. Horodecki, 
e-print arXiv:0805.3060 (2008).
%%%%%%%%%%%%%%%%%%%%%%%%%%%%%%

\bibitem{Dillenschneider:08} R. Dillenschneider, Phys. Rev. B {\bf 78}, 224413 (2008).

\bibitem{Henderson:01} L. Henderson and V. Vedral, J. Phys. A {\bf 34}, 6899 (2001).

\bibitem{Syljuasen:03} O. F. Sylju\aa sen, Phys. Rev. A {\bf 68}, 060301(R) (2003).

%%%%%%%%%%% V2: Ref. included
\bibitem{Osterloh:06} A. Osterloh, G. Palacios, and S. Montangero, 
Phys. Rev. Lett. {\bf 97}, 257201 (2006).
%%%%%%%%%%%%%%%%%%%%%%%%%%%%%%

\bibitem{Oliveira:08} T. R. de Oliveira, G. Rigolin, M. C. de Oliveira, and E. Miranda, 
Phys. Rev. A {\bf 77}, 032325 (2008).

\bibitem{Yang:66} C. N. Yang and C. P. Yang, Phys. Rev. {\bf 150}, 321 (1966); {\it ibid.} 
{\bf 150}, 327 (1966).

\bibitem{Pfeuty:70} P. Pfeuty, Ann. Phys. {\bf 57}, 79 (1970).

\bibitem{Hellmann:37} H. Hellmann, 
{\it Die Einf\"uhrung in die Quantenchemie} (Deuticke, Leipzig, 1937).

\bibitem{Feynman:39} R. P. Feynman, Phys. Rev. {\bf 56}, 340 (1939).

\bibitem{Gu:03} S.-J. Gu, H.-Q. Lin, and Y.-Q. Li, 
Phys. Rev. A {\bf 68}, 042330 (2003).

\bibitem{Gu:04} S.-J. Gu, S.-S. Deng, Y.-Q. Li, and H.-Q. Lin, 
Phys. Rev. Lett. {\bf 93}, 086402 (2004).

\bibitem{Franca:06} V. V. Fran\c{c}a and K. Capelle, Phys. Rev. A {\bf 74}, 042325 (2006).

\bibitem{Bose:02} I. Bose and E. Chattopadhyay, Phys. Rev. A {\bf 66}, 062320 (2002).

\bibitem{Alcaraz:03} F. C. Alcaraz, A. Saguia, and M. S. Sarandy, 
Phys. Rev. A {\bf 70}, 032333 (2004).

\bibitem{Wootters:98}
{W. K. Wootters}, Phys. Rev. Lett. {\bf 80},  2245  (1998).

\bibitem{Vidal:02a}
{G. Vidal, R. F. Werner}, Phys. Rev. A {\bf 65},  032314  (2002).

\bibitem{Yang:05} M.-F. Yang, Phys. Rev. A \textbf{71}, 030302(R) (2005).

\bibitem{LMG:65} H. J. Lipkin, N. Meshkov, and A. J. Glick, Nucl. Phys. {\bf 62}, 188 (1965).

\bibitem{Ring:book} P. Ring and P. Schuck, {\it The Nuclear Many-Body Problem}, 
Springer-Verlag, New York, 1980.

%%%%%%%% V2: 2 Refs. included
\bibitem{Ribeiro:07} P. Ribeiro, J. Vidal, and R. Mosseri, Phys. Rev. Lett. {\bf 99}, 050402 (2007).

\bibitem{Ribeiro:08} P. Ribeiro, J. Vidal, and R. Mosseri, Phys. Rev. E {\bf 78}, 021106 (2008).
%%%%%%%%%%%%%%%%%%%%%%%%%%%%%%

%%%%%%%% V2: Ref. replaced
\bibitem{Vidal:04} J. Vidal, G. Palacios, and R. Mosseri, Phys. Rev. A {\bf 69}, 022107 (2004).
%%%%%%%%%%%%%%%%%%%%%%%%%%

\bibitem{Wu:06} L.-A. Wu, M. S. Sarandy, D. A. Lidar, and L. J. Sham, 
Phys. Rev. A {\bf 74}, 052335 (2006). 

\bibitem{Ma:09} J. Ma and X. Wang, Phys. Rev. A {\bf 80}, 012318 (2009).

%%%%%%%%% V2: Refs included
\bibitem{Shabani:09} A. Shabani and D. A. Lidar, Phys. Rev. Lett. {\bf 102}, 100402 (2009) 

\bibitem{Werlang:09} T. Werlang, S. Souza, F. F. Fanchini, and C. J. Villas-Boas, e-print arXiv:0905.3376 (2009).
%%%%%%%%%%%%%%%%%%%%%%%%%%%%%%%%%%%%%%%%%

%%%%%%%%% V2: Ref. updated.
\bibitem{Maziero:09} J. Maziero, L. C. C\'eleri, R. M. Serra, and V. Vedral, e-print arXiv:0905.3396 (2009).
%%%%%%%%%%%%%%%%%%%%%%%%%%%%%


\end{thebibliography}
\end{document}